\documentclass[aip, amsmath, amssymb,reprint]{revtex4-2}

\usepackage[utf8]{inputenc}
\usepackage{graphicx}
\usepackage{dcolumn}
\usepackage{bm}
\usepackage[utf8]{inputenc}
\usepackage[T1]{fontenc}
\usepackage{mathptmx}
\usepackage{gensymb}
\usepackage{upgreek}
\usepackage{xcolor}
\usepackage{textcomp}
\usepackage{hyperref}
\hypersetup{colorlinks=true,linkcolor=red,citecolor=blue,filecolor=magenta,urlcolor=magenta}

\usepackage{titlesec}
\titleformat{\section}[hang]{\small\bfseries\sffamily}{\thesection.}{0.5em}{\MakeUppercase}
\titlespacing{\section}{0pc}{1pc}{0.2pc}

\begin{document}
\title{Incorporation of erbium ions into thin-film lithium niobate integrated photonics}

\author{Sihao Wang}
\thanks{These two authors contributed equally}
\affiliation{Department of Electrical Engineering, Yale University, New Haven, CT 06511, USA}
\author{Likai Yang}
\thanks{These two authors contributed equally}
\affiliation{Department of Electrical Engineering, Yale University, New Haven, CT 06511, USA}
\author{Risheng Cheng}
\affiliation{Department of Electrical Engineering, Yale University, New Haven, CT 06511, USA}
\author{Yuntao Xu}
\affiliation{Department of Electrical Engineering, Yale University, New Haven, CT 06511, USA}
\author{Mohan Shen}
\affiliation{Department of Electrical Engineering, Yale University, New Haven, CT 06511, USA}
\author{Rufus L. Cone}
\affiliation{Department of Physics, Montana State University, Bozeman, MT 59717, USA}
\author{Charles W. Thiel}
\affiliation{Department of Physics, Montana State University, Bozeman, MT 59717, USA}
\author{Hong X. Tang}
\email{hong.tang@yale.edu}
\affiliation{Department of Electrical Engineering, Yale University, New Haven, CT 06511, USA}

\date{\today}

\begin{abstract}

As an active material with favorable linear and nonlinear optical properties, thin-film lithium niobate has demonstrated its potential in integrated photonics. Integration with rare-earth ions, which are promising candidates for quantum memories and transducers, will enrich the system with new applications in quantum information processing. Here, we investigate the optical properties at 1.5 micron wavelengths of rare-earth ions (Er$^{3+}$) implanted in thin-film lithium niobate waveguides and micro-ring resonators. Optical quality factors near a million after post annealing show that ion implantation damage can be successfully repaired. The transition linewidth and fluorescence lifetime of erbium ions are characterized, revealing values comparable to bulk-doped crystals. The ion-cavity coupling is observed through a Purcell enhanced fluorescence, from which a Purcell factor of ~3.8 is extracted. This platform is compatible with top-down lithography processes and leads to a scalable path for controlling spin-photon interfaces in photonic circuits. 
\end{abstract}
\maketitle

As an important material in modern photonics, lithium niobate (LN) displays favorable piezoelectric, electro-optic, optical, photoelastic and photorefractive properties \cite{weis1985lithium}. It is widely used for electro-optic modulators, frequency doublers, optical parametric oscillators and Q-switches for lasers. However, LN had been limited to bulk crystal components in all of these applications until a recent breakthrough in LN thin-film nanofabrication technology \cite{zhang2017monolithic} made compact and dense photonic integrated circuits possible. Subsequent works on high-performance electro-optical modulators \cite{wang2018integrated, he2019high}, ultra-efficient second harmonic generation \cite{lu2019periodically, chen2019ultra} and microwave-optical transduction \cite{shao2019microwave} have stimulated intense interest and promise for integrated photonics. This advance in LN thin-film nanofabrication technology also raises the interest in incorporating rare-earth ions (REIs) into patterned LN waveguides and micro-cavities for scalable photonic integrated circuits with added functionalities enabled by the REIs. 

REIs are well known for their applications in nonlinear optics such as lasers and amplifiers\cite{thiel2011rare} due to their stable optical transitions, high fluorescence quantum efficiencies, and long population lifetimes. Their narrow homogeneous linewidths\cite{equall1994ultraslow, bottger2009effects}, which allow the burning of ultra-narrow spectral holes, also find themselves useful in photonic signal processing\cite{babbitt14} as well as frequency stabilization \cite{thiel2011rare,strickland2000laser,sellin2001laser}, medical imaging\cite{li2008pulsed} and optical filtering\cite{thiel2011rare,cone2012rare}. Over the past decades, REIs have emerged as a promising candidate for quantum information processing\cite{wesenberg2007scalable, simon10review, tittle10review} thanks to the weak coupling of their 4f electrons to the environment\cite{macfarlane87review} and long coherent spin states \cite{zhong2015optically}. REIs have been used to demonstrate quantum memory protocols for quantum networks \cite{saglamyurek2011broadband}, light-matter interactions\cite{clausen2011quantum} and quantum-state teleportation\cite{bussieres2014quantum}. Among the several well-studied REIs, erbium (Er) has received much attention due to its optical transitions in the telecommunications band, avoiding the need for frequency conversion and reducing the complexity of the system. There are already past efforts in incorporating erbium ions into various hosts for integrated photonics such as bulk LN\cite{jiang2019rare}, yttrium orthosilicate (YSO) crystals\cite{miyazono2017coupling},  and silicon nitride\cite{gong2010linewidth}. However, there is still a knowledge gap in combining erbium ions with LN thin-film structures, which could open a vital avenue to harness the benefit of both systems. 

In this Letter, we investigate the Er doping of LN photonic devices through ion implantation. We observe recovery of the optical performance of the LN thin-film nanostructures after post-annealing that repairs the lattice damage caused by implantation. The transition linewidth and fluorescence lifetime of the implanted ions inside optical waveguides are then characterized. The enhanced interaction of erbium ions with micro-cavities, which can be utilized to overcome the small oscillator strength of REIs, is demonstrated through the Purcell effect.

\begin{figure}[h]
\includegraphics[width=0.45\textwidth]{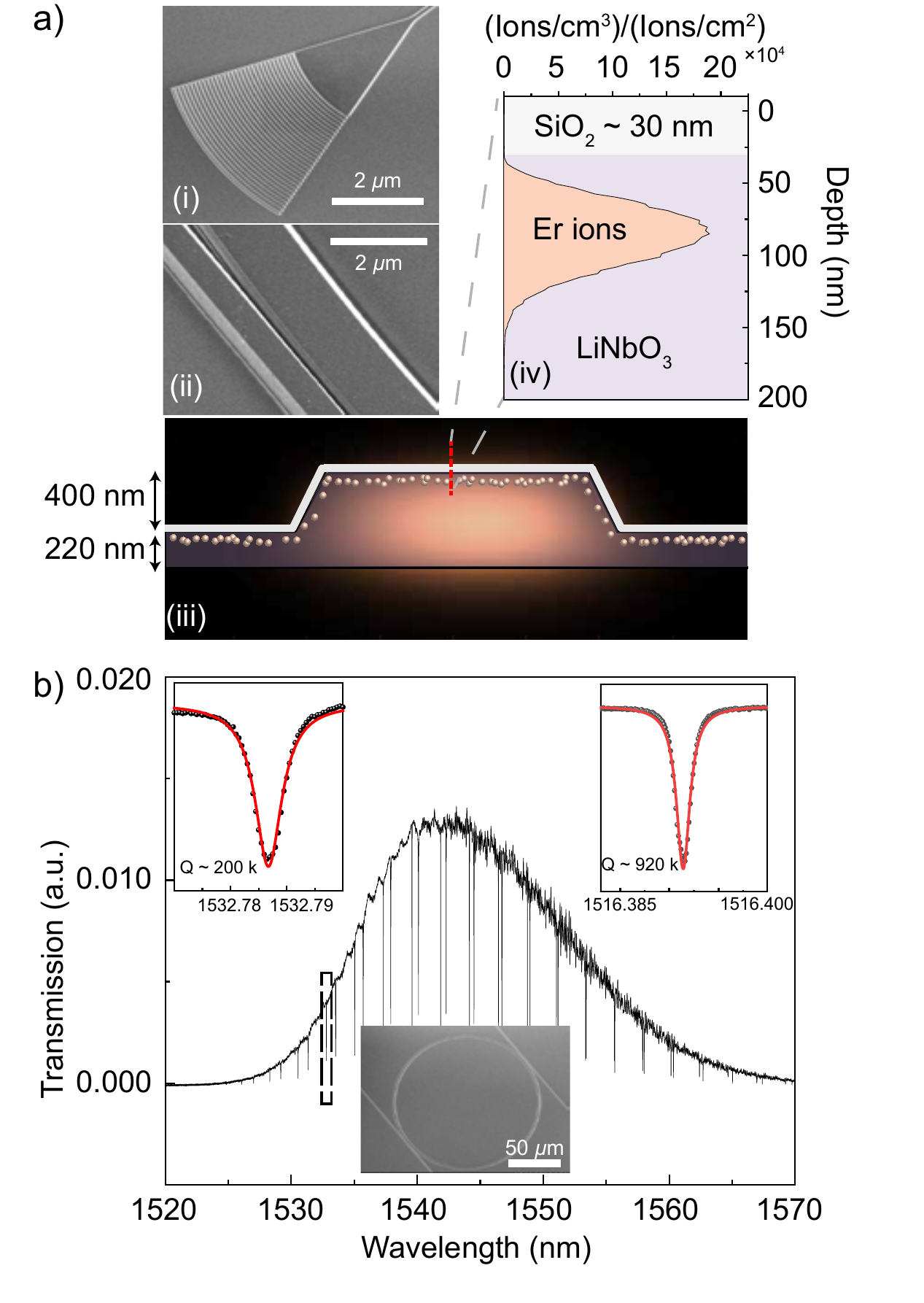}
\caption{\label{fig:1} \textbf{a)i} SEM image of a grating coupler patterned in LN. \textbf{ii} SEM image of the coupling region of a LN ring resonator. \textbf{iii} Schematic cross section of a ring resonator and the fundamental TE mode. The LN thin film is half-etched with a 60\,degree vertical side wall. \textbf{iv} SRIM simulation of the Er implantation depth distribution under a 350\,kV acceleration voltage and a flux fluence of 1.14$\times$10$^{14}$\,ions/cm$^2$. \textbf{b)} The transmission measurement of a ring resonator. The two groups of modes correspond to the fundamental and first-order TE modes. The top-left inset is a resonance with a loaded Q of 200\,k around 1532\,nm indicated by the black-dashed box. The top-right inset is a resonance at 1516\,nm in another resonator, showing a loaded Q of nearly a million. The bottom inset is a SEM image of a typical ring resonator with bus waveguides.}
\end{figure}

Commercial (NanoLN) thin-film wafers of 600\,nm-thick z-cut lithium niobate on insulator (LNOI) are chosen for device fabrication and subsequent ion implantation. We fabricate micro-ring resonators and centimeter-long waveguides by patterning hydrogen silsesquioxane (HSQ) on LN thin film with electron beam lithography (EBL). Ring radius is designed to be 70\,$\mu$m. The ring and the waveguide widths are 1.8\,$\mu$m and 800\,nm respectively. The coupling gap between the ring and the bus waveguide is varied from 400 nm to 1000 nm to identify the critical coupling condition. Grating couplers optimized for the transverse electric (TE) mode at 1532\,nm are used to efficiently screen many devices. The structures are then half-etched through reactive ion etching (RIE) with argon plasma. Fig.~\ref{fig:1}a)i shows the scanning electron microscope (SEM) image of a grating coupler and Fig.~\ref{fig:1}a)ii shows the SEM image of the coupling region of a ring resonator. Fig.~\ref{fig:1}a)iii shows a schematic of the cross section of a ring resonator with a 60 degree etched side wall. The TE fundamental mode is well confined inside the half-etched ring. The ring resonators exhibit an average Q of 500\,k after post fabrication annealing at 500\,\textdegree{}C for 5 hours.

Before erbium implantation, 30\,nm-thick SiO$_2$  is deposited on the chip by plasma-enhanced chemical vapor deposition (PECVD). The erbium ions are doped with an implantation energy of 350\,keV and a flux fluence of 1.14$\times$10$^{14}$\,ions/cm$^2$. The stopping and range of ions in matter (SRIM)\cite{ziegler10SRIM} simulation (Fig.~\ref{fig:1}a)iv) estimates that the peak ion density is around 80 nm deep from the top surface with the tail of the distribution extending to the SiO$_2$/LiNbO$_3$ interface. This corresponds to an average implantation depth of 50\,nm into the LN thin film.

Post-implantation annealing is carried out at various temperatures to repair the lattice damage from implantation. A temperature of 350\,\textdegree{}C is sufficient to recover the transmission, but the maximum Q is not achieved until 550\,\textdegree{}C. A typical grating coupler transmission measurement is shown in Fig.~\ref{fig:1}b), in which fundamental and first-order TE modes of the resonator can be observed. The bottom inset of Fig.~\ref{fig:1}b) shows a typical ring resonator with bus waveguides. A resonance with a Q of around 200\,k at around 1532\,nm in the device used for later experiments is shown in the inset at the top-left corner. It is worth noting that several high-Q resonances could be found in other ring resonators, one of which had a Q of nearly a million at 1516\,nm, as shown in the inset at the top-right corner. The average Q of the devices after post-implantation annealing at 550\,\textdegree{}C is ~500\,k, indicating that a full recovery of optical performance of the nanostructure is possible.

\begin{figure}[h]
\includegraphics[width=0.45\textwidth]{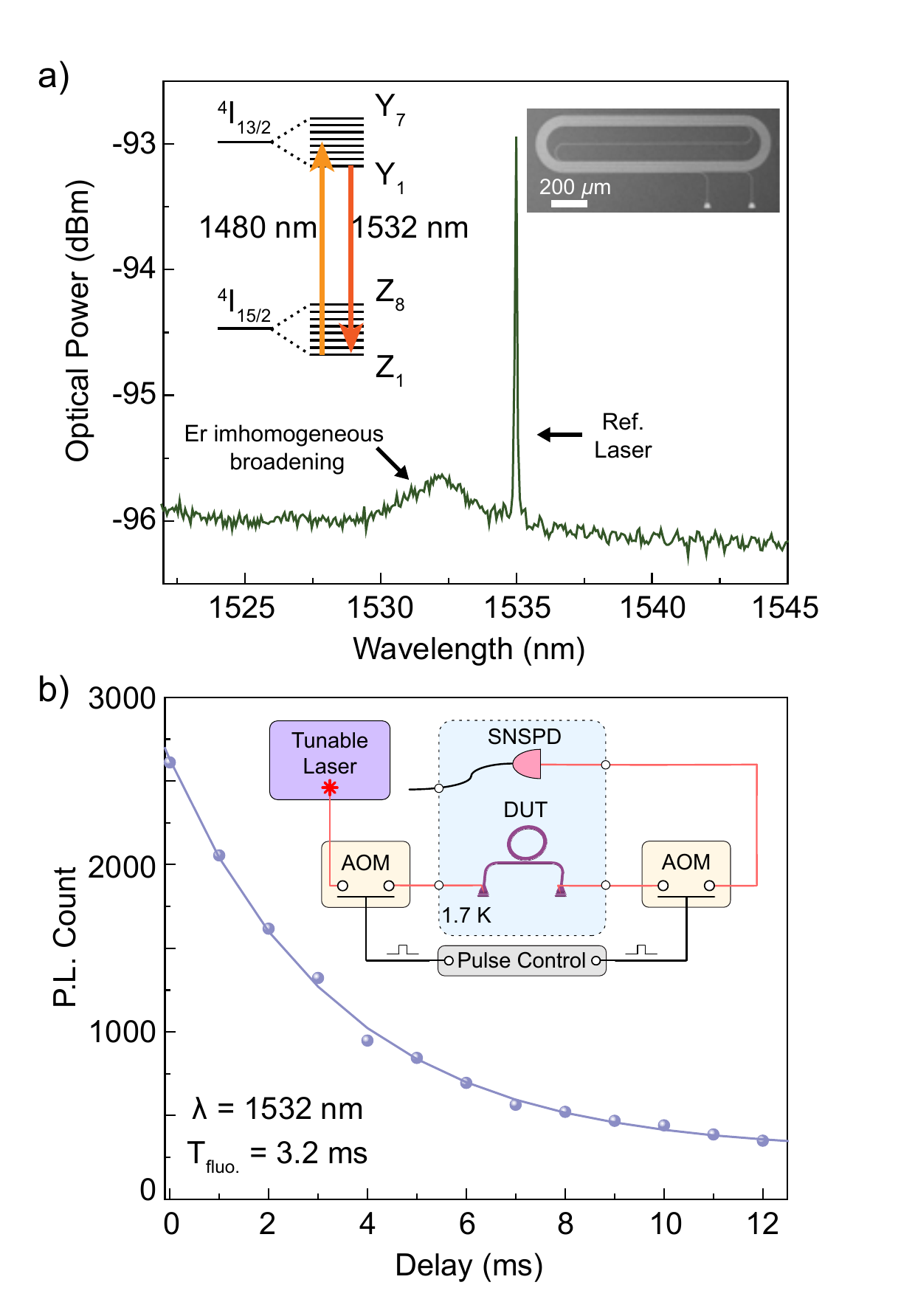}
\caption{\label{fig:2} \textbf{a)} Non-resonant fluorescence measurement at room temperature with an optical pump at 1480\,nm. A tunable filter is applied to collect the spectral response of the fluoresence around 1532\,nm. A reference laser is combined via a wavelength division multiplexer to calibrate the tunable filter. The small peak near 1532\,nm is due to the Er$^{3+}$ fluorescence. The sharp peak at 1535\,nm is due to the reference laser used for wavelength calibration. The relevant energy levels of Er:LN are shown in the top-left inset. The crystal field splits the $^{2S+1}L_J$ states of the free Er$^{3+}$ ions into to $J+1/2$ Kramer doublets that can be further split with an applied magnetic field. The top-right inset is a SEM image of a typical 4.7\,cm long waveguide. \textbf{b)} Resonant fluorescence measurement at 1.7\,K with an optical pump at 1532\,nm. The fluorescence signal is gated and delayed for lifetime measurements. The solid line is a single exponential fit with a time constant of ~3.2\,ms. The inset shows the experimental schematic of the measurement.}
\end{figure}

Fig.~\ref{fig:2}a) shows a room temperature fluorescence measurement to map the inhomogeneous broadening of erbium ions. The experiment is carried out on a 4.7\,cm-long waveguide (top-right inset of Fig~\ref{fig:2}a)). A strong optical pump at 1480\,nm is used to excite the Er$^{3+}$ ions and the spectral response is mapped with a tunable filter. A reference laser is introduced to calibrate the tunable filter. 

The crystal structure of LN is described by space group $C_{3v}^6$ where Li$^{+}$ and Nb$^{3+}$ ions sit along the trigonal axis. When the Er$^{3+}$ ions sit at the Li$^+$ site, their point group is $C_3$.\cite{nolte97resite} This crystal field splits the Er$^{3+}$ ground electronic state ($^4I_{15/2}$) into 8 Kramer doublets labeled as $Z_{1-8}$ and the Er excited electronic state ($^4I_{13/2}$) into 7 Kramer doublets labeled as $Y_{1-7}$. The optical pump at 1480\,nm coincides with a strong absorption peak of the $Z_1-Y_6$ transition\cite{gruber2004modeling}. Fast non-radiative decay\cite{bottger2006spectroscopy} first relaxes the state to $Y_1$ and then optically relaxes to the ground Stark state $Z_1$ by emiting a photon at 1532\,nm. This process is illustrated in the top-left inset of Fig.~\ref{fig:2}a). The Er$^{3+}$ fluorescence signal as well as the signature of the reference laser are shown clearly in Fig.~\ref{fig:2}a). The room temperature Er$^{3+}$ fluorescence peak is centered at around 1532\,nm, and a Lorentz fit yields a linewidth of 250\,GHz.  This linewidth is comparable to the literature value of 180\,GHz for the low-temperature inhomogeneous broadening in bulk-doped crystals \cite{thiel2011rare}. The small increase is likely due to additional homogeneous broadening from phonon scattering at room temperature, although there may also be a contribution from crystal strain\cite{dibos2018atomic} caused by bonding the LN thin film to the oxide substrate. The device is then cooled down to 1.7\,K, where a resonant excitation measurement is done to extract the population lifetime of the Er$^{3+}$ ions. The inset of Fig.~\ref{fig:2}b) shows the experiment schematic. Two accousto-optic modulators (AOM) are placed before and after the device for gating to resonantly pump and measure the fluorescence at 1532\,nm using a continuous-wave tunable laser and a fiber-coupled superconducting nanowire single-photon detector (SNSPD)\cite{cheng2016self}. A single exponential fit of the fluorescence decay allows us to extract a lifetime of 3.2\,ms.  This is somewhat longer than the excited-state lifetime of 2.0\,ms measured for a bulk-doped crystal by spectral hole burning\cite{thiel10erln}. Fluorescence lifetimes for bulk-doped crystals are usually significantly longer than the excited-state lifetime due to the strong fluorescence absorption and re-emission that occurs for this transition\cite{nunez96}, which likely also causes the longer lifetime observed here.

\begin{figure}[h]
\includegraphics[width=0.45\textwidth]{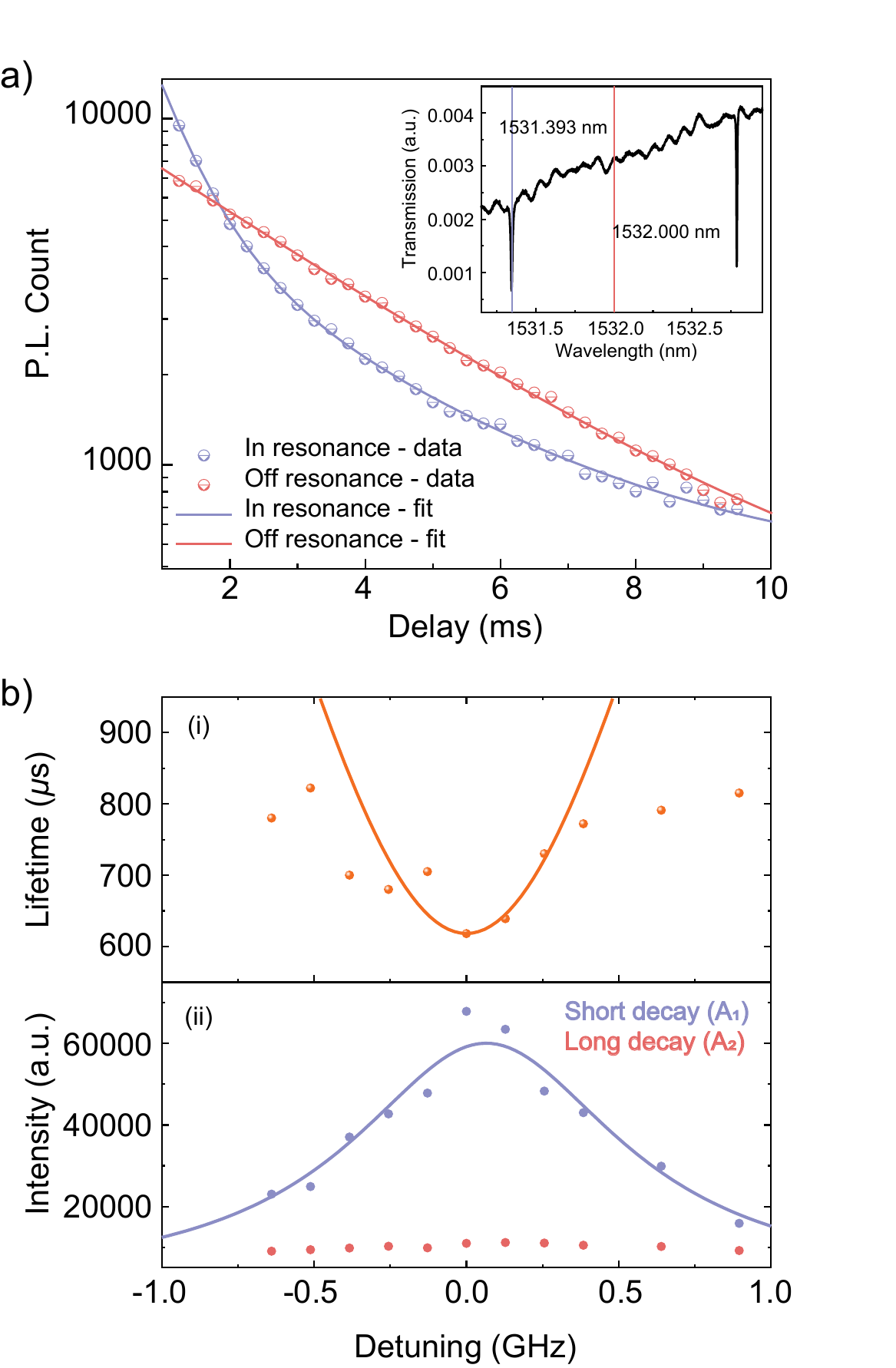}
\caption{\label{fig:3} \textbf{a)} Fluorescence decay measurement on a semi-log scale for on-resonance pumping at 1531.393\,nm (blue) and off-resonance pumping at 1532.000\,nm (orange). The data from the off-resonance pump is fitted with a single exponential function, while those from the on-resonance pump is fitted with a double exponential function where the long decay time constant is fixed at the constant value obtained from the single-exponential fit of the fluorescence from a waveguide. The inset shows the transmission of the ring resonator near 1532\,nm. The blue (orange) line indicates the on(off)-resonance pump wavelength. \textbf{b)i} Fluorescence lifetime from the short decay of the double exponential fit in 3\textbf{a)}. The solid line is the theoretical prediction based on the Purcell enhancement due to the ring resonance. \textbf{ii} The fluorescence intensities from the double exponential fit in 3\textbf{a)}. The short decay (blue) represents the fluorescence from the ring resonator and the long decay (red) represents the fluorescence from the bus waveguide. The solid line is the theoretical prediction based on the cavity enhancement.   }
\end{figure}

To demonstrate the ion-cavity coupling, we focus on a resonance near 1532\,nm in a ring resonator and extract the Purcell-enhanced fluorescence decay rate. The fluorescence decay when pumping on- and off-resonance are compared in Fig.~\ref{fig:3}a). Due to the interaction with the cavity field, the fluorescence lifetime of an ion in the ring resonator $(T_{\mathrm{ring}})$ should be shortened compared to the value in the bus waveguides outside of the cavity $(T_{\mathrm{wg}})$, according to\cite{bienfait2016controlling}
\begin{eqnarray}
\frac{1}{T_{\mathrm{ring}}} = \frac{1}{T_{\mathrm{wg}}} + \frac{\kappa g^2}{(\kappa/2)^2 + \delta^2},
\label{eq:one}
\end{eqnarray}
where $\kappa=\omega_0/Q$ is the decay rate of the ring resonator and $\delta=\omega-\omega_0$ is the detuning from the resonance frequency $(\omega_0)$. The coupling rate between the ion and resonator can be written as
\begin{eqnarray}
g=\frac{\mu}{n}\sqrt{\frac{\omega}{2\epsilon_0\hbar V}}\frac{|E(r)|}{|E_{\mathrm{peak}}|},
\label{eq:two}
\end{eqnarray}
where $\mu$, $n=2.2$ \cite{schlarb93} and $V=\frac{\int dV \epsilon(r) |E(r)|^2}{\mathrm{max}(\epsilon(r)|E(r)|^2)}$ are the dipole moment, refractive index and resonator mode volume. The coupling also depends on the position of the ion, which is quantified by the spatial overlap with the cavity mode $|E(r)|/|E_{\mathrm{peak}}|$. By substituting Eq.~\ref{eq:two} into Eq.~\ref{eq:one}, we arrive at the formula for the Purcell factor at zero detuning
\begin{eqnarray}
P = \frac{T_{\mathrm{wg}}}{T_{\mathrm{ring}}}-1=\frac{3}{4 \pi^2}\left( \frac{\lambda_0}{n} \right)^3 \frac{Q}{V} \frac{|E(r)|^2}{|E_{\mathrm{peak}}|^2}.
\label{eq:three}
\end{eqnarray}

To quantify the coupling rate and the Purcell factor, we first fit the fluorescence measurement off-resonance from the cavity with a single exponential decay function
\begin{eqnarray}
f_{\mathrm{off}} = A_0 \mathrm{Exp}\left( - \frac{t}{T_{\mathrm{wg}}}   \right),
\label{eq:four}
\end{eqnarray}
where $T_{\mathrm{wg}}\approx3$\,ms is obtained. For the on-resonance case, however, there are two groups of ions contributing to the fluorescence, e.g. one in the ring resonator, and the other in the bus waveguide. For simplicity, we approximate the influence of the cavity field with a Purcell enhanced lifetime averaged over all of the ions at different locations. Then the overall fluorescence from on-resonance pumping should be fitted with a double exponential decay function, fixing the long time constant at $T_{\mathrm{wg}}$,
\begin{eqnarray}
f_{\mathrm{in}} = A_1 \mathrm{Exp}\left( - \frac{t}{T_{\mathrm{ring}}}   \right) + A_2 \mathrm{Exp}\left( - \frac{t}{T_{\mathrm{wg}}}   \right)
\label{eq:five}.
\end{eqnarray}
As shown in Fig.~\ref{fig:3}a), the experimental data are well approximated by this double exponential decay function, giving $T_{\mathrm{ring}}=618\,\mu$s. The average Purcell factor is then calculated to be $P_{\mathrm{avg}}=3.8$, and the coupling rate $g_{\mathrm{avg}}=(2\pi)\;597$\,kHz.

The measured value of Purcell enhancement can be theoretically modeled by Eq.~\ref{eq:three}. For the selected resonance at 1531.393\,nm, we measure $Q=217$\,k. From the simulation results of the field distribution, we calculate the mode volume $V=223$\,${\mu}$m$^3$, and an average value of field overlap $|E(r)|^2/|E_{peak}|^2 = 1/8$ for the simulated implantation depth of 50\,nm. This results in an overall Purcell enhancement $P=3.1$, which is close to the experimental value of 3.8. We note that further improvement of the Purcell enhancement requires a resonance with a higher $Q$, a smaller mode volume $V$ and a better spatial overlap $|E(r)|^2/|E_{peak}|^2$. For the current batch of devices, the enhancement could be about 5 times higher if the high-Q resonance was centered around the Er$^{3+}$ transition wavelength of 1532\,nm. Another factor of 8 improvement would also be possible if the Er$^{3+}$ ions are incorporated into the center of the resonator. 

We also examine the fluorescence decay when the pumping frequency is detuned from the resonance. Fig.~\ref{fig:3}b)i shows the measured lifetime of the ions in the ring resonator, extracted from the double exponential fit, as well as the theoretical curve calculated based on Eq.~\ref{eq:one}. The corresponding fluorescence intensities from the ring and the bus waveguide portions are plotted in Fig.~\ref{fig:3}b)ii. It is clear that the fluorescence from the waveguide ($A_2$) is insensitive to the detuning from the resonance, while the fluorescence from the ring resonator ($A_1$) peaks at zero detuning. Likewise, we theoretically model the enhanced fluorescence signal from the ions in the ring resonator ($A_1$) by a Lorentzian curve $A_{1}^{\mathrm{theo}} = \frac{C}{(\kappa/2)^2+(\delta)^2}$ with $C$ being a fitting parameter. The model matches the data well in Fig.~\ref{fig:3}b)ii, showing a clear resonance-enhanced fluorescence decay.

In conclusion, we introduce Er-implanted LN thin-film nanostructures and investigate their optical properties. We show that the cavity properties in the doped device could be recovered by post-implantation annealing with a temperature of up to 550\,\textdegree{}C, obtaining resonances with a loaded Q of near a million after annealing. The transition linewidth and population lifetime of the erbium ions are extracted in waveguides and resonant cavities through fluorescence measurements. We show that the optical emission of ions in the ring resonator is cavity-enhanced and that the experimental data match theoretical models well, with an average Purcell enhancement of ~3.8. This work demonstrates the possibility to fabricate a compact and scalable platform for spin-photon interfaces with LN-based integrated photonics.

\textit{Note added}: In preparing this manuscript, the authors noticed that a related pre-print discussing a different rare-earth element (Tm) in bulk-doped LN thin-film waveguides\cite{dutta2019integrated}.

This work is supported by Department of Energy, Office of Basic Energy Sciences, Division of Materials Sciences and Engineering under Grant DE-SC0019406. The authors would like to thank Dr. Yong Sun, Sean Rinehart, Kelly Woods, and Dr. Michael Rooks for their assistance provided in the device fabrication. The fabrication of the devices was done at the Yale School of Engineering \& Applied Science (SEAS) Cleanroom and the Yale Institute for Nanoscience and Quantum Engineering (YINQE).\\

\noindent\textbf{REFERENCES}


%

\end{document}